\documentclass[namedreferences]{SolarPhysics}
\usepackage[optionalrh]{spr-sola-addons} % For Solar Physics
\usepackage{graphicx}        % For eps figures, newer & more powerfull
\usepackage{color}           % For color text: \color command
\usepackage{url}             % For breaking URLs easily trough lines
\usepackage{lineno}

\newcommand{\apj}{    {\it Astrophys. J.}}

\linenumbers*[1]

\def\referee#1{$\blacktriangleright${#1}$\blacktriangleleft$}

\def\referee#1{{#1}}

\begin{document}

\begin{article}

\begin{opening}

\title{Can We Improve the Preprocessing of Photospheric Vector Magnetograms
by the Inclusion of Chromospheric Observations?}

\author{T.~\surname{Wiegelmann}$^{1}$\sep
        J.K.~\surname{Thalmann}$^{1}$\sep
        C.J.~\surname{Schrijver}$^{2}$\sep
        M.L.~\surname{DeRosa}$^{2}$\sep
        T.R.~\surname{Metcalf}$^{3}$
        \footnote{Unfortunately our colleague, co-author, and friend Tom Metcalf deceased
  before the final manuscript was finished. We continued our joint work in his
  memory and would like to remember him here.}
       }
\runningauthor{Wiegelmann et al.}
\runningtitle{H$\alpha$ Preprocessing}
\date{DOI: 10.1007/s11207-008-9130-y \\
Bibliographic Code: 2008SoPh..247..249W }
   \institute{$^{1}$ Max-Planck-Institut f\"ur Sonnensystemforschung,
Max-Planck-Strasse 2, 37191 Katlenburg-Lindau, Germany email:
\url{wiegelmann@mps.mpg.de} email: \url{thalmann@mps.mpg.de}\\ $^{2}$ Lockheed
Martin Advanced Technology Center, Dept. ADBS. Bldg. 252, 3251 Hanover St.,
Palo Alto, CA 94304, USA email: \url{schryver@lmsal.com} email:
\url{derosa@lmsal.com} \\ $^{3}$ Northwest Research Associates, Colorado
Research Associates Division, 3380 Mitchell Ln., Boulder, CO 90301 USA }

\begin{abstract}
The solar magnetic field is key to understanding the physical processes in the
solar atmosphere. Nonlinear force-free codes have been shown to be
useful in extrapolating the coronal field upward from underlying vector
boundary data.  However, we can only measure the magnetic field vector
routinely with high accuracy in the photosphere, and unfortunately these data
do not fulfill the force-free condition. We must therefore apply some
transformations to these data before nonlinear force-free extrapolation codes
can be self-consistently applied.  To this end, we have developed a
minimization procedure that yields a more chromosphere-like field, using the
measured photospheric field vectors as input.  The procedure includes
force-free consistency integrals, spatial smoothing, and --- newly included in
the version presented here --- an improved match to the field direction as
inferred from fibrils as can be observed in, {\it e.g.}, chromospheric H$\alpha$
images.
We test the procedure using a model active-region field that included buoyancy
forces at the photospheric level.  The proposed preprocessing method allows us
to approximate the chromospheric vector field to within a few degrees and the
free energy in the coronal field to within one percent.
\end{abstract}
\end{opening}
%-------------------------------------------------

\section{Introduction}
The solar interior, photosphere and atmosphere are coupled by magnetic fields.
It is therefore important to gain insights about the magnetic field
structure in all layers of the Sun and solar atmosphere. Direct and
accurate measurements of the magnetic field vector are typically
carried out only on the photosphere. Although measurements in higher
layers are available for a few individual cases, {\it e.g.} in the chromosphere by
%\inlinecite{solanki:etal03}
Solanki {\it et al.} (2003)
and in the corona by \inlinecite{lin:etal04}, the
line-of-sight integrated character of such chromospheric and coronal magnetic
field measurements complicates their interpretation \cite{Kramar:etal06}.
Knowledge of the magnetic field in the corona is essential, however, to
understand basic physical processes such as the onset of flares,
coronal mass ejections and eruptive prominences.

Inferences of the coronal magnetic field can be obtained by
extrapolating measurements of the photospheric magnetic field vector
({\it e.g.} observed by {\it Hinode}/SOT, SOLIS or the upcoming SDO/HMI
instruments) into the corona. Because the magnetic pressure
dominates the plasma pressure in active-region coronae, making the plasma
$\beta$ low, [See work by \inlinecite{gary01} and
\inlinecite{schrijver:etal05}, which discuss the plasma beta over
active regions and over the quiet Sun, respectively], these
extrapolations neglect non-magnetic forces and assume the coronal magnetic
field ${\bf B}$ to be force-free, such that it obeys:

\begin{eqnarray}
\nabla \cdot {\bf B} &=& 0 \label{divB}, \\
(\nabla \times {\bf B}) \times {\bf B} &=& 0 \label{jxb}.
\end{eqnarray}

Equation (\ref{jxb}) implies that the electric current density ${\mu_0 \bf
j}=\nabla \times {\bf B}$ is parallel to the magnetic field ${\bf
B}$. Starting more than a quarter century ago \cite{sakurai81}, different
mathematical methods and numerical implementations have been developed to
solve the nonlinear force-free equations (\ref{divB}) and (\ref{jxb}) for
the solar case.  See, for example,
\cite{sakurai89,amari:etal97,wiegelmann07a} for review papers and
\cite{schrijver:etal06,metcalf:etal07} for evaluations of the
performance of corresponding computer programs with model data. The codes use
the magnetic field vector (or quantities derived from the magnetic field
vector) on the bottom boundary of a computational domain as
input. One would like to prescribe the measured photospheric data as the
bottom boundary of nonlinear force-free fields (NLFFF) codes, but there is a
problem: the observed photospheric magnetic field is usually not
force-free. The relatively high plasma $\beta$ in the photosphere
means that non-magnetic forces cannot be neglected there and that such
photospheric magnetic field data are not consistent with well known force-free
compatibility conditions defined in \cite{aly89}.  Recently,
\inlinecite{wiegelmann:etal06} developed a scheme that mitigates this
problem, in which the inconsistent and noisy photospheric vector magnetograms
used as bottom boundary conditions are preprocessed in order to remove net
magnetic forces and torques and to smooth out small-scale noise-like magnetic
structures.  The resulting magnetic field data are sufficiently
force-free and smooth for use with extrapolation codes, but also are found to
bear a high resemblance to chromospheric vector magnetic field data.  This
leads us to the question whether we can constrain the preprocessing tool
further by taking direct chromospheric observations, such as H$\alpha$ images,
into consideration.  We will investigate this topic in the present
work.
\section{A Short Review About Consistency Criteria for Force-free Coronal Extrapolations}
In this section, we briefly discuss the criteria on the photospheric boundary
data that are required for consistency with a force-free extrapolation of the
overlying coronal magnetic field. \inlinecite{molodensky69},
\inlinecite{molodensky74}, \inlinecite{aly89}, and \inlinecite{sakurai89} show
how moments of the Lorentz force, integrated over a volume of interest, define
constraints on the closed surface bounding this volume.
As explained in detail in \inlinecite{molodensky74} the sense of these relations
is that on average a force-free field cannot exert pressure on the boundary or shear stresses
along axes lying in the boundary.
For the coronal magnetic field extrapolation calculations discussed here, a
localized region of interest, such as an active region, is typically selected
for analysis. The extrapolation algorithms applied to the coronal volume
overlying such localized regions of interest require boundary conditions, and,
except at the lower (photospheric) boundary, these boundary conditions are
usually chosen to be consistent with potential fields and thus do not possess
magnetic forces or torques. In these cases, the consistency criteria reduce
to conditions on the lower boundary only:

\begin{enumerate}
\item On average force-free fields cannot exert pressure on the boundary

 \begin{eqnarray}
 F_1=\int_{S} B_x B_z \;dx\,dy &=& 0,
 \label{prepro1} \\
 F_2=\int_{S} B_y B_z \;dx\,dy &=& 0 \\
 F_3=\int_{S} (B_x^2 + B_y^2) \; dx\,dy  - \int_{S} B_z^2 \; dx\,dy &=&0.
 \label{prepro2}
 \end{eqnarray}

\item On average force-free fields cannot create shear stresses along axes lying
in the boundary

 \begin{eqnarray}
 T_1=\int_{S} x \; (B_x^2 + B_y^2) \; dx\,dy  -\int_{S} x \; B_z^2 \; dx\,dy &=&0,
 \label{prepro3} \\
 T_2=\int_{S} y \; (B_x^2 + B_y^2) \; dx\,dy - \int_{S} y \; B_z^2 \; dx\,dy &=&0,
 \label{prepro4} \\
 T_3=\int_{S} y \; B_x B_z \; dx\,dy - \int_{S} x \; B_y B_z \; dx\,dy &=&0.
 \label{prepro5}
 \end{eqnarray}
\end{enumerate}
These relations must be fulfilled in order to be suitable boundary conditions
for a nonlinear force-free coronal magnetic field extrapolation.
We define dimensionless numbers,

\begin{equation}
\epsilon_{\mbox{force}}=\frac{\rm |F_1| + |F_2| + |F_3| }
{\int_{S} (B_x^2+B_y^2+B_z^2) \;dx\,dy},
\label{eps_force}
\end{equation}
\begin{equation}
\epsilon_{\mbox{torque}}=\frac{\rm |T_1| + |T_2| + |T_3|}
{\int_{S} \sqrt{x^2+y^2} \; (B_x^2+B_y^2+B_z^2) \; dx\,dy}.
\label{eps_torque}
\end{equation}
in order to evaluate how well these criteria are met. Ideally, it is
necessary for $\epsilon_{\mbox{force}}=\epsilon_{\mbox{torque}}=0$ for a
force-free coronal magnetic field to exist.
%In practice, however, the
%preprocessing algorithms minimize these quantities as much as possible. We
%use the values of $\epsilon_{\mbox{force}}$ and $\epsilon_{\mbox{torque}}$ in
%Section~\ref{section:results} to evaluate the efficacy of the preprocessing
%schemes presented in this study.

\inlinecite{aly89} pointed out that the magnetic field is probably not
force-free in the photosphere, where ${\bf B}$ is measured because the plasma
$\beta$ in the photosphere is of the order of unity and pressure gradient and
gravity forces are not negligible. The integral relations
(\ref{prepro1})-(\ref{prepro5}) are not satisfied in this case in the
photosphere and the measured photospheric field is not a suitable boundary
condition for a force-free extrapolation. Investigations by
%\inlinecite{metcalf:etal95}
Metcalf {\it et al.} (1995)
revealed that the solar magnetic field is not
force-free in the photosphere, but becomes force-free about $400 {\rm km}$
above the photosphere. The problem has been addressed also by
\inlinecite{gary01} who pointed out that care has to be taken when
extrapolating the coronal magnetic field as a force-free field from
photospheric measurements, because the force-free low corona is sandwiched
between two regions (photosphere and higher corona) with a plasma $\beta
\approx 1$, where the force-free assumption might break down. An additional
problem is that measurements of the photospheric magnetic vector field contain
inconsistencies and noise. In particular the components of ${\bf B}$
transverse to the line of sight, as measured by current vector magnetographs,
are more uncertain than the line-of-sight component. As measurements in
higher layers of the solar atmosphere (where the magnetic field is force-free)
are not routinely available, we have to deal with the problem of inconsistent
(with the force-free assumption as defined by Equations (\ref{prepro1})--(\ref{prepro5}))
photospheric measurements. A routine which uses measured photospheric vector
magnetograms to find suitable boundary conditions for a nonlinear force-free
coronal magnetic field extrapolation, dubbed ``preprocessing'', has been
developed by \inlinecite{wiegelmann:etal06}.

\section{Preprocessing} \label{sec:preprocessing}
\subsection{Classical Preprocessing}
%The integral relations
% (\ref{prepro1})-(\ref{prepro5}) have been used to define
% a 2D functional of quadratic form:
The preprocessing scheme of \inlinecite{wiegelmann:etal06} involves minimizing
a two-dimensional functional of quadratic form similar to the following:

 \begin{equation}
  L_{\rm prep} = \mu_1 L_1 + \mu_2 L_2 + \mu_3 L_3 + \mu_4 L_4 + \mu_5 L_5,
  \label{defLprep}
 \end{equation}
where

\begin{eqnarray}
%Force
L_1 &=& \left[ \left(\sum_p B_x B_z \right)^2
              +\left(\sum_p B_y B_z \right)^2
  +\left(\sum_p B_z^2-B_x^2-B_y^2 \right)^2 \right],
  \label{defL_1} \\
%Torque
L_2 &=& \left[ \left(\sum_p x  \left(B_z^2-B_x^2-B_y^2 \right) \right)^2
              +\left(\sum_p y  \left(B_z^2-B_x^2-B_y^2 \right) \right)^2
        \right. \nonumber \\ & & \left. \hspace*{0.8em}
              +\left(\sum_p y B_x B_z -x B_y B_z \right)^2
        \right], \\
%Data
L_3 &=& \left[ \sum_p \left(B_x-B_{xobs} \right)^2
              +\sum_p \left(B_y-B_{yobs} \right)^2 \right. \nonumber \\
 &&      \left. +\sum_p \left(B_z-B_{zobs} \right)^2 \right],
 \label{defL_3} \\
%Smoothing
L_4 &=& \left[ \sum_p \left(\Delta B_x \right)^2
                     +\left(\Delta B_y \right)^2
                     +\left(\Delta B_z \right)^2
        \right].
\label{defL_4}
\end{eqnarray}

The surface integrals
as defined in Equations (\ref{prepro1})--(\ref{prepro5})  are here
replaced by a summation $\sum_p$ over all grid
nodes $p$ of the bottom surface grid.  We normalize the magnetic field
strength with the average magnetic field on the photosphere and the length
scale with the size of the magnetogram.  Each constraint $L_{\rm n}$ is weighted by
a yet undetermined factor $\mu_{\rm n}$. The first term ($n$=1) corresponds to the
force-balance conditions (\ref{prepro1})-(\ref{prepro2}), the next ($n$=2) to
the torque-free condition (\ref{prepro3})-(\ref{prepro5}). The following term
($n$=3) contains the difference of the optimized boundary condition with the
measured photospheric data and the next term ($n$=4) controls the
smoothing. The 2D-Laplace operator is designated by $\Delta$ and the
differentiation in the smoothing term is achieved by the usual 5-point
stencil.  The last term ($n=5$) has not been used in preprocessing so far and
will be introduced in the next section.
The aim of the preprocessing procedure is to minimize $L_{\rm prep}$ so that
all terms $L_{\rm n}$ if possible are made small simultaneously. This minimization
procedure provides us iterative equations for $B_x, B_y, B_z$ (see
\inlinecite{wiegelmann:etal06} for details). As result of the preprocessing we
get a data set which is consistent with the assumption of a force-free
magnetic field in the corona but also as close as possible to the measured
data within the noise level.

Nonlinear force-free extrapolation codes can be applied
only to low plasma $\beta$ regions, where the force-free assumption is justified.
This is known not to be the case in the photosphere, but is mostly true for the upper
chromosphere and for the corona in quiescent conditions.  The preprocessing scheme
as used until now  modifies observed photospheric vector magnetograms with the aim
of approximating the magnetic field vector at the bottom of the force-free domain,
{\it i.e.}, at a height that we assume to be located in the middle to upper chromosphere.
In this study, we investigate whether the use of chromospheric fibril observations
as an additional constraint in the preprocessing can bring the resulting field into
even better agreement with the expected chromospheric vector field.

We discuss this idea in the next section.
\subsection{H$\alpha$-Preprocessing}
 The idea is to specify another term ($\mu_5 \, L_5$) in
 Equation (\ref{defLprep}) which measures how well the preprocessed magnetic field
 is aligned with fibrils seen in H$\alpha$.
 As a first step we have to extract the directions of the fibrils, say
 $H_x$ and $H_y$ out of the H$\alpha$ images, where ${\bf H}$ is a unit
 vector tangent $(|{\bf H}| = 1)$ to the chromospheric fibrils projected onto the solar
 photosphere (representing the field direction with a 180-degree ambiguity).
 For simplicity one might rebin $H_x$ and $H_y$ to the same resolution as
 the vector magnetogram.
 In regions where we cannot identify  clear filamentary structures in
 the images we set $H_x=H_y=0$. These regions are only affected by the other,
 classical terms of the preprocessing functional (\ref{defLprep}).
 The angle of the projected magnetic field vector on the xy-plane
 with the H$\alpha$ image is

 \begin{equation}
 %\sin(\phi)=\frac{{\bf B} \times {\bf H}}{|B||H|},
 \sin(\phi) = \frac{|{\bf B}_{\parallel} \times {\bf H}|}{|{\bf B}_{\parallel}| |{\bf H}|}
 \label{angle_phi}
 \end{equation}
where ${\bf B}_{\parallel}=(B_x,B_y)$ is the projection of the magnetic field vector
in the xy-plane and ${\bf H}=(H_x,H_y)$ are the directions of the chromospheric
H$\alpha$ fibrils.
The preprocessing aims for deriving the magnetic field vector
on the bottom boundary of the force-free domain, which is located in the chromosphere.
The chromospheric magnetic field is certainly a priori unknown and as initial condition
for the preprocessing routine we take ${\bf B}_{\parallel}$  from the photospheric
vector magnetogram.

 We define the functional:

 \begin{equation}
 L_5=\sum_p w (B_x H_y - B_y H_x)^2
  = \sum_p w  {\bf B}_{\parallel}^2 \sin^2(\phi).
 \label{defL5}
 \end{equation}

 Please note that the term $B_x H_y - B_y H_x$ in Equation (\ref{defL5}) weights
 the angle with the magnetic field strength, because it is in particular
 important to minimize the angle in strong field regions. The
 space dependent function $w=w(x,y)$ is not a priori related to the
 magnetic field strength. $w$ can be specified in order to
 indicate the confidence level of the fibril direction-finding algorithm
 (see {\it e.g.}, \inlinecite{inhester:etal07} for the description of a corresponding
 feature recognition tool). For the application to observational data $w$
 will be (with appropriate normalization) provided by this tool. It is
 likely, however, that the direction of the H$\alpha$ fibrils can be identified more
 accurately in strong magnetic field regions, but this is not an a priori assumption.
 In Section \ref{opti_halpha} we investigate the influence of different assumptions
 for $w$.

 We take the functional derivative of $L_5$

 \begin{equation}
 \frac{d L_5}{dt}=2(B_x H_y - B_y H_x) \; (H_y \frac{d B_x}{dt}-H_x \frac{d
 B_y}{dt}).
 \end{equation}

 For a sufficiently small time step $dt$ we get a decreasing $L_5$ with
 the iteration equations

 \begin{eqnarray}
 \frac{d B_x}{dt} & = & -2 w \, \mu_5 (B_x H_y - B_y H_x) \, H_y, \\
 \frac{d B_y}{dt} & = & 2 w \, \mu_5 (B_x H_y - B_y H_x) \, H_x.
 \end{eqnarray}

 The aim of our procedure is to make all terms in functional (\ref{defLprep})
 small simultaneously. There are obvious contradictions between some
 of the $L_{\rm n}$ terms, such as between the $n=3$ (photospheric data) and $n=4$
 (smoothing) terms.
 An important task is to find suitable values for the five parameters
 $\mu_{\rm n}$ which control the relative weighting of the terms in Equation
 (\ref{defLprep}).
 The absolute values do not matter; only the
 relative weightings are important. We typically
 give all integral relations of the force and torque conditions
 (\ref{prepro1})-(\ref{prepro5}) the same weighting (unity). To fulfill these
 consistency integrals is essential in order to find
 suitable boundary conditions for a nonlinear force-free extrapolation.
 In principle it would be possible to examine different values for the force-free term
 $\mu_1$ and torque-free term $\mu_2$ -or even to give six different weightings
 for the six integral relations- but giving all integrals the same weighting seems
 to be a reasonable choice. The torque integrals depend on the choice of
 the length scale $D$ and giving the same weighting to all integrals requires
 $\mu_2=\frac{\mu_1}{D^2}$. For the length scale normalization used here
 $(D=1)$ this leads to $\mu_1=\mu_2$.

 We will test our
 newly developed method with the help of a model active region in the next
 section.
\section{Tests}
\begin{figure}
\includegraphics[width=14cm]{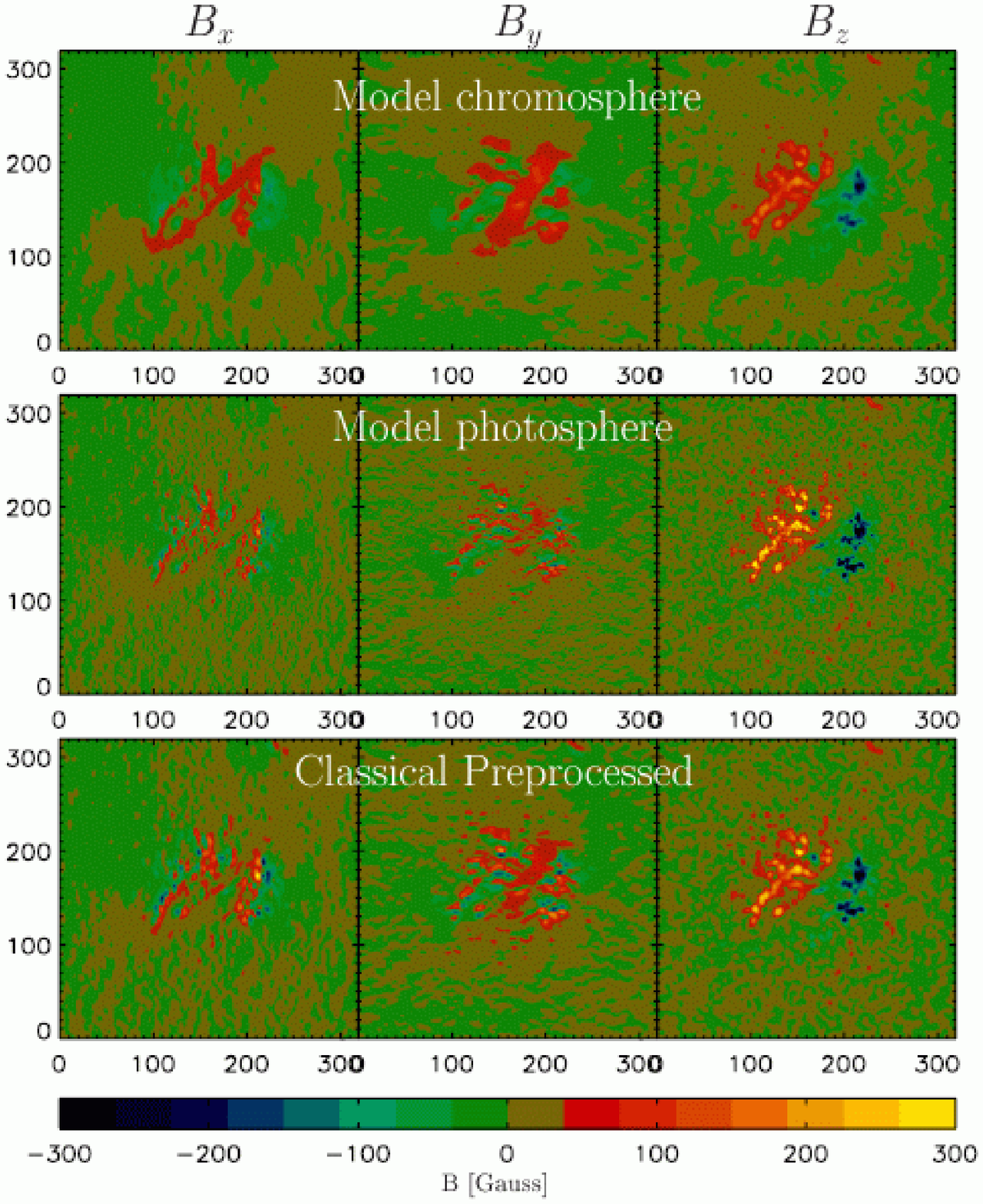}
\caption{Top: Model-chromospheric magnetic field located in the $z=2$ layer,
Center: Model-photospheric magnetic field, Bottom:
Model-photospheric magnetic field after classical preprocessing with
$\mu_3=0.025, \, \mu_4=0.155$.}
\label{magnetograms1}
\end{figure}
\begin{figure}
\includegraphics[width=14cm]{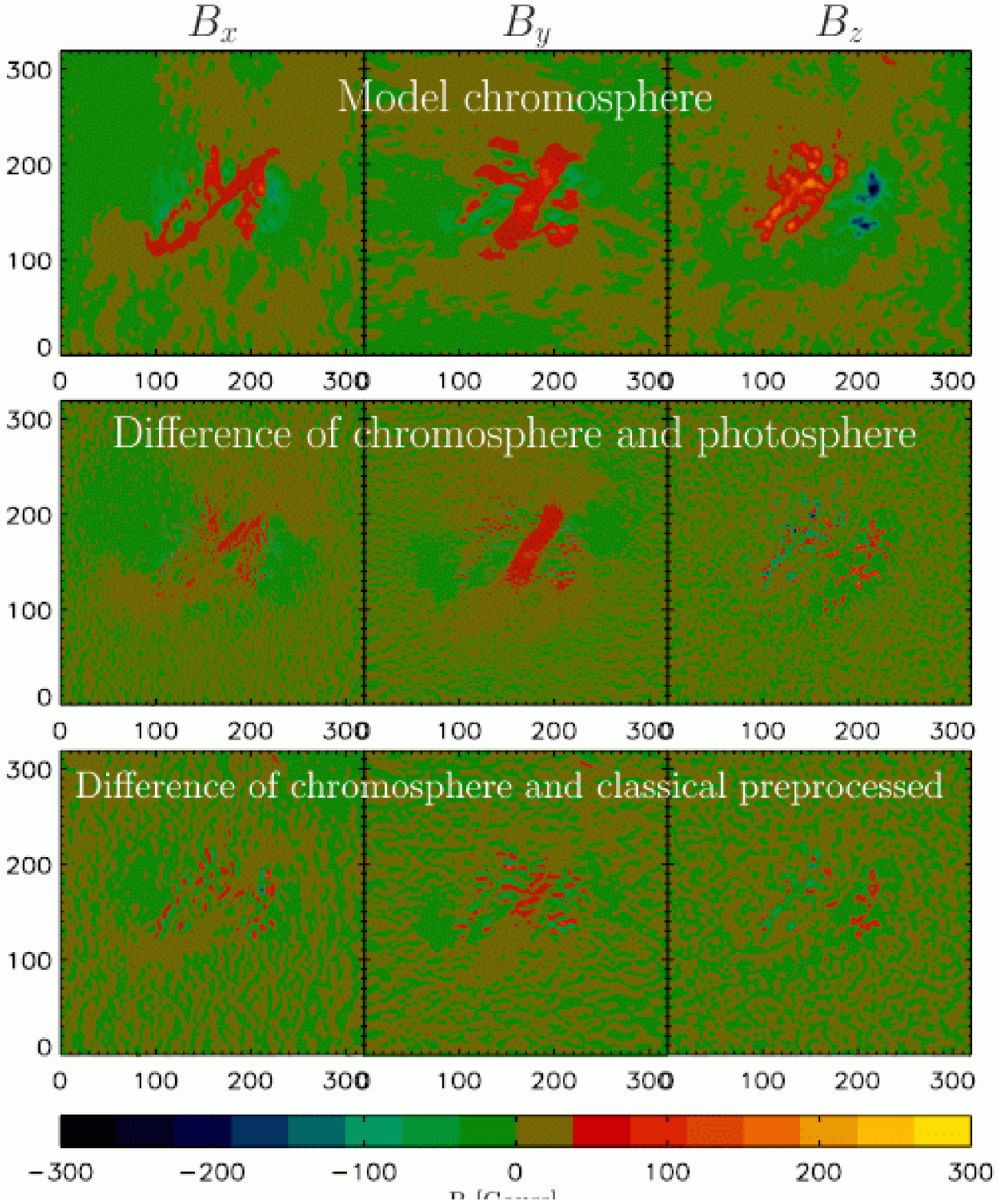}
\caption{Top: Model-chromospheric magnetic field,
Center: Difference between the chromospheric and photospheric
model vector field,
Bottom: Difference between the chromospheric and classical preprocessed
photospheric field.}
\label{magnetograms1a}
\end{figure}
\subsection{An Active Region Model for Testing the New Method}
We test our extended preprocessing routine with the help of an active region
model recently developed by
%\inlinecite{vanballegooijen:etal07}.
van Ballegooijen {\it et al.} (2007)
In this model
line-of-sight photospheric measurements from SOHO/MDI have been used to
compute a potential field.  A twisted flux rope was then inserted
into the volume, after which the whole system was relaxed towards a nonlinear
force-free state with the magnetofrictional method described in
\inlinecite{vanballegooijen04}. The
%\inlinecite{vanballegooijen:etal07}
van Ballegooijen {\it et al.} (2007)
model
is force-free throughout the entire computational domain, except
within two gridpoints of the bottom boundary.  Hereafter, we refer to the
bottom of the force-free layer as the ``model chromosphere'' (see the top
panel of Figure \ref{magnetograms1}).  On the bottom boundary (see the central
panel of Figure \ref{magnetograms1}), hereafter referred to as the ``model
photosphere'', the model contains significant non-magnetic forces and the
force-free consistency criteria (\ref{prepro1})-(\ref{prepro5}) are not
satisfied. These forces take the form of vertical buoyancy forces directed
upward, and have been introduced by %\inlinecite{vanballegooijen:etal07}
van Ballegooijen {\it et al.} (2007)
to mimic
the effect of a reduced gas pressure in photospheric flux tubes. The nature of
these forces is therefore expected to be similar to those observed on the real
Sun. For a more detailed discussion we refer to
%\inlinecite{metcalf:etal07}
Metcalf {\it et al.} (2007).
Both the chromospheric ($\textbf{B}_{\rm ch}$) as well as the photospheric
magnetic field vector ($\textbf{B}_{\rm ph }$) from the
%\inlinecite{vanballegooijen:etal07}
van Ballegooijen {\it et al.} (2007)
model have been used to test four sophisticated
nonlinear force-free extrapolation codes in a blind algorithm test by
%\inlinecite{metcalf:etal07}
Metcalf {\it et al.} (2007).
\footnote{Previously, the NLFFF codes have been
intensively tested and evaluated with the \cite{low:etal90} semi-analytic
equilibria \cite{schrijver:etal06}.}  The codes computed nonlinear force-free
codes in a $320 \times 320 \times 256$ box, which is about at the upper limit
current codes can handle on workstations.  We briefly summarize the results of
%\inlinecite{metcalf:etal07}
Metcalf {\it et al.} (2007)
as:
 \begin{itemize}
 \item NLFFF-extrapolations from model-chromospheric data recover
       the original reference field with high accuracy.
 \item When the extrapolations are applied to the model-photospheric
       data, the reference field is not well recovered.
 \item Preprocessing of the model-photospheric data to remove net
       forces and torques improves the result, but the resulting accuracy was
       lower than for extrapolations from the model-chromospheric data.
 \end{itemize}
The poor performance of extrapolations using the unprocessed
model-photospheric data is related to their inconsistency with respect to the
force-free conditions (\ref{prepro1})-(\ref{prepro5}).  The central
panel of Figure  \ref{magnetograms1} shows the photospheric magnetic field and
the central panel of Figure \ref{magnetograms1a} illustrates the difference
between the model-chromospheric and model-photospheric fields. It is evident
that there are remarkable differences in all components of the magnetic field
vector.  For real data we usually cannot measure the chromospheric magnetic
field vector directly (which was possible for \cite{vanballegooijen:etal07}
model data) and we have to apply preprocessing before using the data as input
for force-free extrapolation codes. Force-free extrapolations using
preprocessed data from the model photosphere (as lower panels of
Figures \ref{magnetograms1} and \ref{magnetograms1a}), while encouraging, were
not completely satisfactory, in light of the results being worse than when the
model-chromospheric data were used as boundary conditions.  In what follows,
we will use an artificial H$\alpha$ image created from the model chromosphere
to test a modified preprocessing scheme, and compare the results to
the classical (original) preprocessing scheme.

We use the model-chromospheric magnetic field $(\textbf{B}_{\rm ch})$ to derive
the direction vectors of the artificial H$\alpha$ images. For the model case
we can simply use the chromospheric model field to specify the direction
vectors $H_x$ and $H_y$, which contain only information regarding the
direction of the horizontal components of the magnetic field
(including a $180^\circ$ ambiguity, but no information about the magnetic
field strength.  For real data this information can be derived from
high-resolution H$\alpha$ images using feature recognition techniques,
{\it e.g.} the ridge detector of
\inlinecite{inhester:etal07}.

\subsection{Optimal Parameter Set for Classical Preprocessing}
\begin{figure}
\includegraphics[width=14cm]{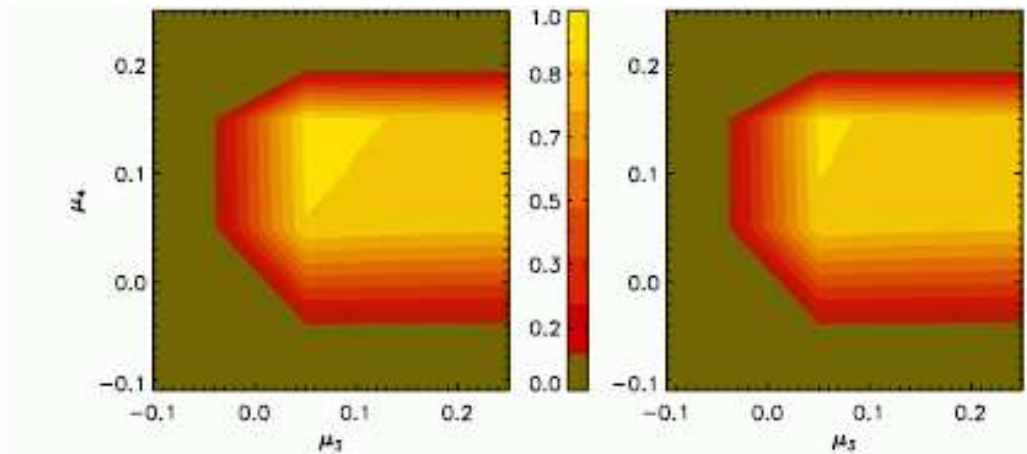}
\caption{Correlation of the preprocessed field (left panel: $B_x$, right
panel: $B_y$) with the model chromosphere in dependence of the preprocessing
parameters $\mu_3$ and $\mu_4$.
We found a maximum correlation at
$\mu_{3} = 0.025$ and $\mu_{4} = 0.155$.}
\label{corr34}
\end{figure}
We tested more than $1000$ possible combinations of $\mu_{3}$ and $\mu_{4}$
using the model-photospheric field as input, and computed the Pearson
correlation coefficient between the preprocessed results and the
model-chromospheric field.  Only $B_x$ and $B_y$ were used in computing the
correlation coefficient, because the correlation of the longitudinal ({\it i.e.},
the line-of-sight) component is in general higher than that of the transverse
components, due to $B_{z}$ not being affected by the ambiguity-problem and the
noise being much lower than in the other directions.

%In the top panels of Figure~\ref{corr34} the correlation is shown for $100$
We computed $100$
combinations of $\mu_{3}~\rm{and}~\mu_{4}$ between $-0.2 \leq \mu_{3}, \mu_{4}
\leq 0.2$ with a step size of $\Delta \mu_{3} = \Delta \mu_{4} =
0.05$. Hereafter a local maximum around $\mu_{3} = 0.05~\rm{and}~\mu_{4} =
0.15$ appeared.
This region was analyzed in more detail by using these two
values as new initial guess. To do this, we tried another $100$ combinations
around this pair with a reduced step size of $\Delta \mu_{3} = \Delta \mu_{4}
= 0.005$ in the positive as well as the negative direction. Then the absolute
maximum of the correlation coefficients for both, $B_{x}$ and $B_{y}$ appeared
at $\mu_{3} = 0.025~\rm{and}~\mu_{4} = 0.155$ (see Figure~\ref{corr34}).
The bottom panel of Figure \ref{magnetograms1} shows the corresponding
preprocessed photospheric magnetic field.

\subsection{Optimal Parameters and Weighting Functions for H$\alpha$ Preprocessing.}
\label{opti_halpha}
%
%%%%%%%%%%%%%%%%%%%%%%%%%%%%%%%%%%%%%%%%%%%%%%%%%%%%%%%%%%%%%%%%%%%%%%%%%%%%%%%%%%%%%%%%%%%%%%%%%%%%
\begin{figure}
\includegraphics[width=14cm]{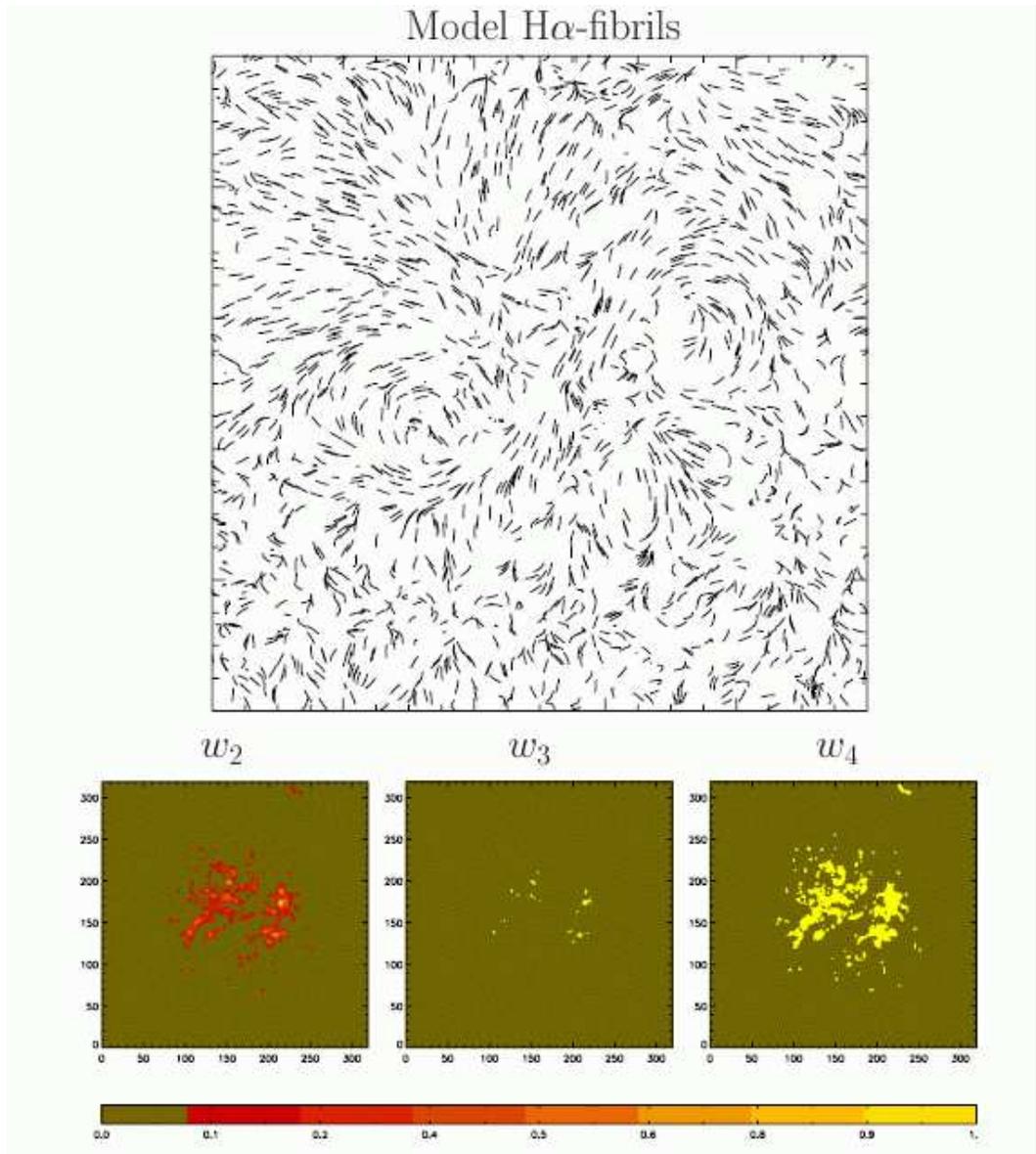}
\caption{Top: H$\alpha$ fibrils identified from the model
chromosphere. The fibrils give us information about the transverse
components ($B_x$ and $B_y$) of the chromospheric magnetic field. The fibrils
contain a $180^{\circ}$ ambiguity and do not provide any information about the
chromospheric magnetic field strength.  The bottom panels show from left to
right the different weighting functions $w_2, w_3, w_4$, respectively.
Regions where $w$ is higher are more important is the
$L_5$-preprocessing-term (\ref{defL5}) which controls the influence of the
H$\alpha$-fibrils.}
\label{fig_halpha}
\end{figure}
In the following we aim to find suitable parameters for including
information from H$\alpha$ images into the preprocessing.

Our main aim is to investigate the effects of additional
chromospheric information. To exclude side effects we therefore
keep the combination of $\mu_1$-$\mu_4$ found in the previous section
to be able to clearly investigate the effect of the additional term $L_5$.
In principle one could
vary all $\mu_{\rm n}$ simultaneously.  We cannot exclude that there might exist a
better combination of $\mu_1$ to $\mu_5$ with better agreement of our preprocessed
field and the model chromospheric field. This is, however, not the aim of this work,
because this is not a suitable way to deal with real data, because there is no model
chromosphere to test the result. It is not possible to provide an optimal parameter set
suitable for all vector magnetographs. The optimal combination has
to be carried out for different instruments separately.
We expect that an optimal parameter set for a certain instrument and
particular region will be also useful for the preprocessing of other regions of
the same kind (say active regions) observed with the same instrument.

We test our methods
with ``model fibrils'' extracted from the model chromosphere shown
in the top panel of Figure \ref{fig_halpha}.  We define $w(x,y)$ used in
Equation (\ref{defL5}) as one of the following:
\begin{enumerate}
\item We assume that at every point of our H$\alpha$ image gives us the exact
orientation of the magnetic field (which is indeed the case, as we calculated
it from the chromospheric model data) and fix our weighting with $w(x,y) = w_1
= 1$.
\item  We assume that the photospheric magnetic field magnitude gives us the importance of the
 H$\alpha$ information at each point and use\\
      \begin{center}
      $w(x,y) = w_2 =  \sqrt{(B_x^2 + B_y^2 + B_z^2)_{\rm ph }}$.
      \end{center}
      We scale $w_2$  to a maximum value of $1$.
      (See Figure  \ref{fig_halpha} bottom left panel.)
\item We do as in the previous case, but assume now, that
 only points in the magnetogram where the field magnitude is greater
 than 50 \% of the maximum
contribute to the H$\alpha$
      preprocessing. So, we define
      \begin{center}
      $w(x,y) = w_3 = \left\{\begin{array}{r} 1 \quad \mbox {\rm{for}} \quad
                                            w_2 \geq 0.5\\ 0 \quad \mbox
                                            {\rm{for}} \quad w_2 < 0.5
                                            \end{array} \right. .$
      \end{center}
       (See Figure  \ref{fig_halpha} bottom center panel.)
\item In our last case we assume in the same way as in the previous one, but
      now  only points in the magnetogram where the field magnitude is greater
 than 10 \% of the maximum
 contribute to the preprocessing. All these grid points are weighted with
      $1$ and the rest with zero. In other words, one defines
      \begin{center}
      $w(x,y) = w_4 = \left\{\begin{array}{r} 1 \quad \mbox {\rm{for}} \quad
                                          w_2 \geq 0.1\\ 0 \quad \mbox
                                          {\rm{for}} \quad w_2 < 0.1
                                          \end{array} \right. .$
      \end{center}
       (See Figure  \ref{fig_halpha} bottom right panel.)
\end{enumerate}
We now figure out the optimal value of $\mu_5$ in Equation (\ref{defLprep}) for the
four different weighting functions $w_1-w_4$.  Initially, we use a step size
of $\Delta \mu_{5} = 0.05$ and then, around the first appearing maximum, we
reduced it to $\Delta \mu_{5} = 0.005$. This is to find a more precise optimal
value of $\mu_5$.  We calculate the Pearson correlation coefficient between
the chromospheric reference field ($\textbf{B}_{\rm ch}$) and the minimum solution
of the preprocessing routine ($\textbf{B}_{pp}$).  This provides us the
optimal values of $\mu_5$ for the different weighting functions, see second
row in Table \ref{tab:testparms2}.
%%%%%%%%%%%%%%%%%%%%%%%%%%%%%%%%%%%%%%%%%%%%%%%%%%%%%%%%%%%%%%%%%%%%%%%%%%%%%%%%%%%%%%%%%%%%%%%%%%%%
\section{Results}
\label{section:results}
\begin{figure}
\includegraphics[width=14cm]{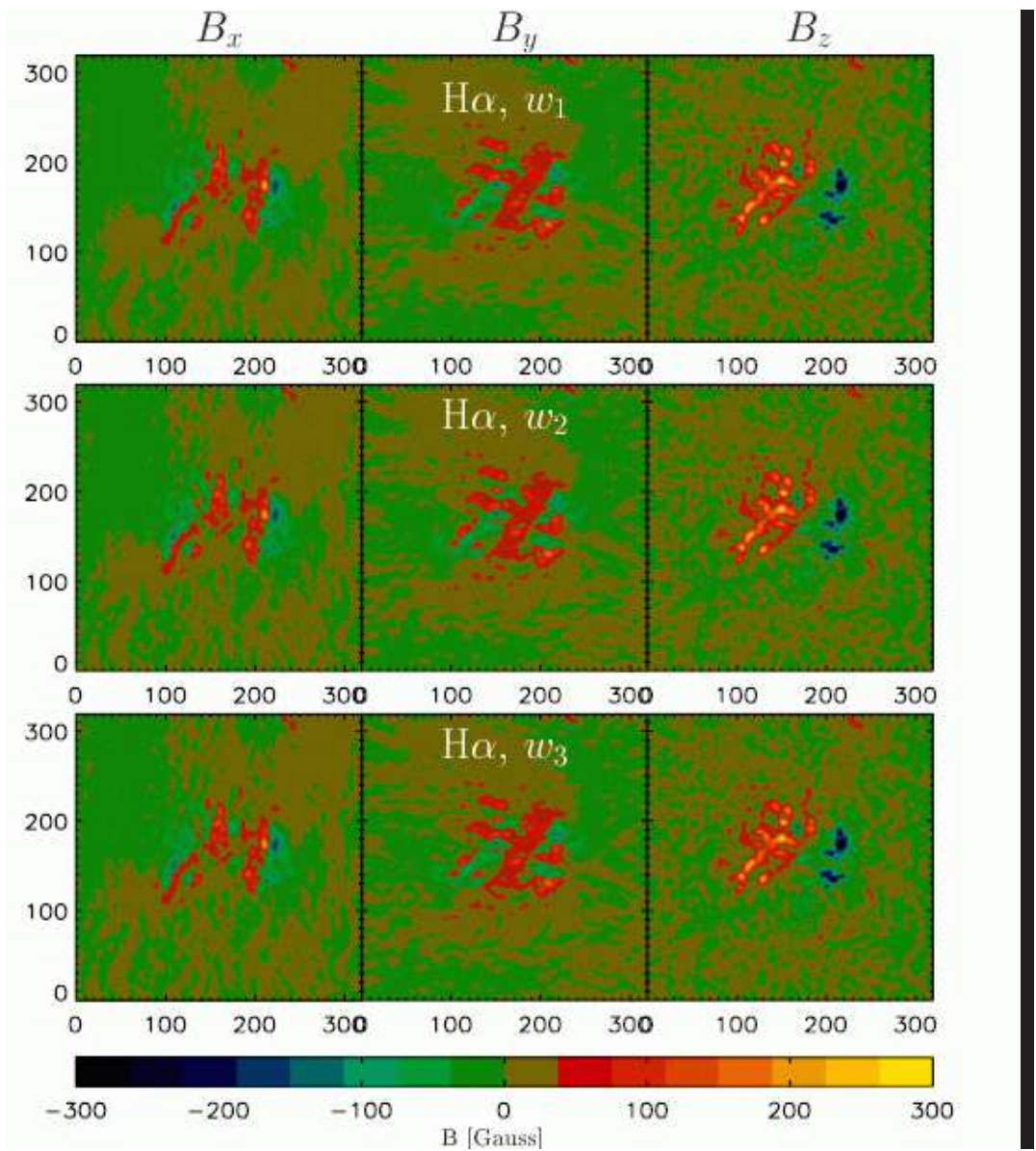}
\caption{Results of H$\alpha$ preprocessing with different weighting
functions.
Top: $w_1$, Center: $w_2$, Bottom: $w_3$, see text.}
\label{magnetograms2}
\end{figure}
\begin{figure}
\includegraphics[width=14cm]{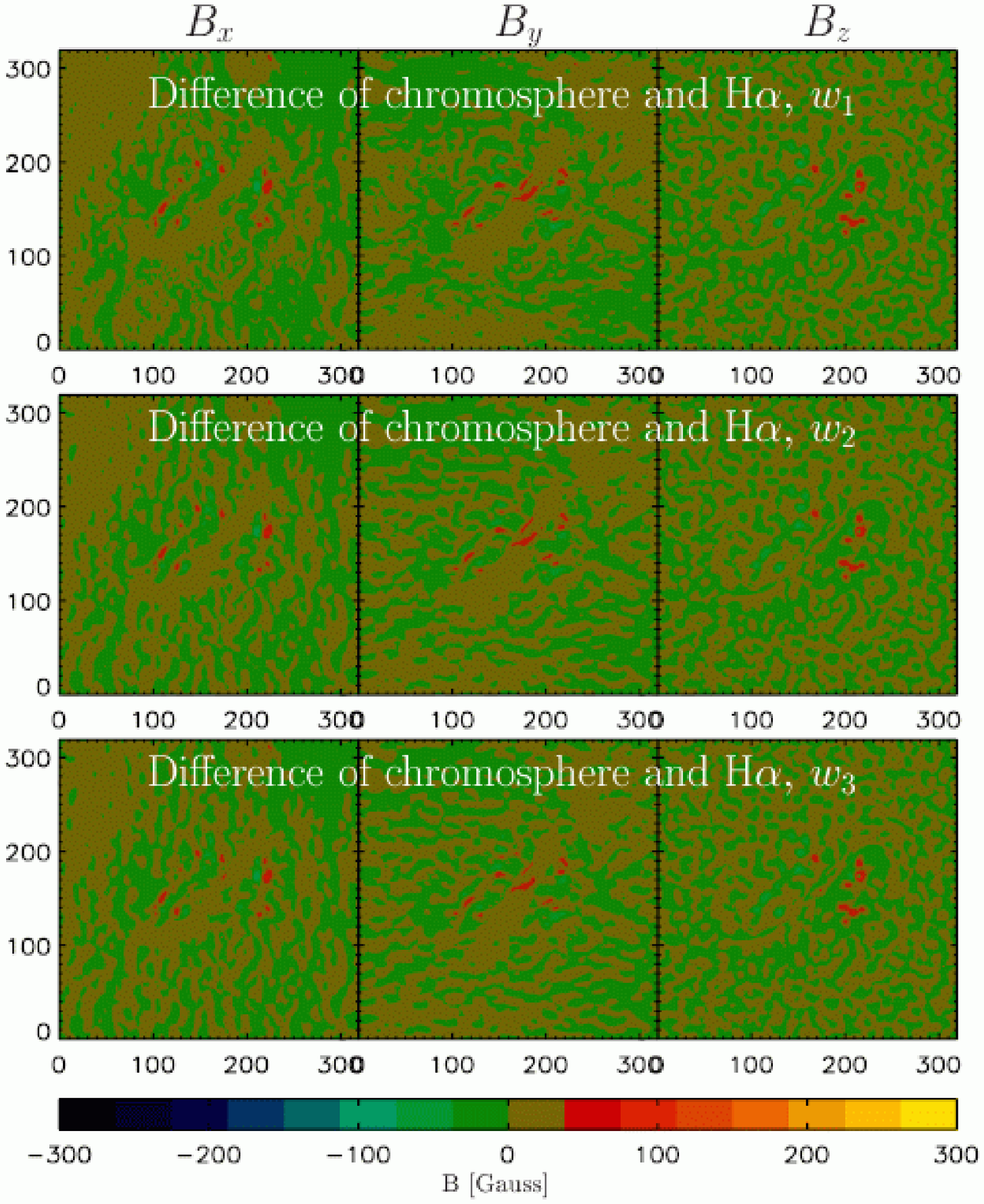}
\caption{Differences of the chromospheric model field (see top panel of
Figure \ref{magnetograms1a} and the H$\alpha$-preprocessed fields as shown
in Figure \ref{magnetograms2}.}
\label{magnetograms2a}
\end{figure}

Table \ref{tab:testparms2} lists some metrics related to the various
preprocessing schemes, including the dimensionless numbers
$\epsilon_{\mbox{force}}$ and $\epsilon_{\mbox{torque}}$ from Equations
(\ref{eps_force}) and (\ref{eps_torque}), the values of the various $L_{\rm n}$ from
Section \ref{sec:preprocessing}, and the averaged angles between the
preprocessing results and the model-chromospheric field.  The first three rows
of the table list the model chromosphere ($B_{\rm ch}$) and photosphere ($B_{\rm ph }$)
data and the classical preprocessing scheme ($B_{\rm cp}$).  When using the
unprocessed model-photospheric data ($B_{\rm ph }$), it is clear that the
force-free consistency criteria (as represented by $L_{12}$,
$\varepsilon_{\rm force}$, and $\varepsilon_{\rm torque}$) are not fulfilled and are
orders of magnitude higher than for the chromospheric data ($B_{\rm ch}$).
Consequently, we cannot expect the extrapolation codes to result in a
meaningful nonlinear force-free field in the corona, as discussed in
%\inlinecite{metcalf:etal07}
Metcalf {\it et al.} (2007).

The remaining rows in Tables \ref{tab:testparms2} and
\ref{tab:extraparms} list the results for the cases where the H$\alpha$
preprocessing was used.  A qualitative comparison of the
H$\alpha$-preprocessed magnetograms (shown in Figure \ref{magnetograms2}) with
the model chromosphere (shown in the top panel of Figure \ref{magnetograms1})
indicates a strong resemblance for all three magnetic field components, but
certainly not a perfect match.  Difference images between the
H$\alpha$-preprocessed magnetograms and the model chromosphere (shown in the
top panel of Figure \ref{magnetograms1}) are present in
Figure \ref{magnetograms2a}.  The resemblance using the H$\alpha$ preprocessing
scheme is much improved when compared to the magnetograms resulting from the
classical preprocessing scheme.

Table \ref{tab:extraparms} displays metrics of the resulting
nonlinear force-free extrapolations using each preprocessing
scheme.\footnote{\referee{For an explanation of the extrapolation method used
to perform the results in Table \ref{tab:extraparms}, see appendix
\ref{appendixA} and references therein.  An explanation of the vector
comparison metrics used in the table is given in Appendix
\ref{appendixB}.}}  As expected, the extrapolation codes perform poorly when
the unprocessed boundary ($B_{\rm ph }$) is used.  In particular, the resulting
magnetic energy $\epsilon_{\rm mag}$ of this case (normalized to the energy of the
reference solution) is only 65\% of the correct answer, making it almost
impossible to estimate the free magnetic energy in the solution available for
release during eruptive processes such as flares and coronal mass ejections.

Taking preprocessing into account (rows 3-7 in both tables) significantly
improves the result.  The force-free consistency criteria
($L_{12},\varepsilon_{\rm force}, \varepsilon_{\rm torque}$) are adequately
fulfilled for all preprocessed cases and are even better (lower values) than
the model chromospheric field. This is naturally, however, because the
preprocessing routine has been developed in particular to derive
force-free-consistent boundary conditions from inconsistent (forced, noisy)
photospheric measurements. The classical preprocessing $(B_{\rm cp})$ has already
reduced the angle to the model H$\alpha$ fibrils (last two columns of Table
\ref{tab:testparms2}) by almost a factor of two, even though no
information about the chromosphere has been used. If we include chromospheric
information, (see Figure \ref{fig_halpha}) in our preprocessing routine
($B_{\rm H\alpha p}$, rows 4-7) the angle of the preprocessed field with the
H$\alpha$ images reduces significantly. The second to last row in Table
\ref{tab:testparms2} contains the average angle and in the last column the
angle has been weighted by the magnetic field, which means that $\phi_{\rm
ave,w}$ measures mainly how well the magnetic field and the chromospheric
fibrils are aligned in regions of a high magnetic field strength. For the
purpose of coronal magnetic field extrapolations the strong field regions are
essential. If we include all information from the H$\alpha$ image, as done in
row 4 for $w_1$ we find that the magnetic field and the fibrils are almost
parallel in the entire region. This is the ideal case, however, as fibrils
have been identified all over the region with the same excellent accuracy. For
observed data it is more likely that the direction of the fibrils
will be identifiable with high accuracy only in bright and
magnetically strong regions. This effect is taken into account in rows 5-7 of
both tables. In the last two rows we take the chromospheric data only into
account where the magnetic field strength is larger than $50\%$ and $10\%$ of
the maximum field strength, respectively. Naturally, the average angle
$\phi_{\rm ave}$ of the chromospheric fibrils with the preprocessed magnetic
field becomes larger than for the ideal case.  We find, however, that the
angle $\phi_{\rm ave,w}$ remains relatively low in strong field regions,
except for the case $w_3$.
%which is however still significantly better than
%classical preprocessing.

We can easily understand that $w_3$ (chromospheric
information ignored where the magnetic field is less than $50 \%$ of its
maximum) provides less accurate results, because the area where chromospheric
data have been taken into account, is only a very small fraction of the entire
region (see Figure \ref{fig_halpha} lower central panel).

Case $w_3$ has few nonzero points. These points are, however, in the
regions with the strongest magnetic field strength. The $L_5$ terms
minimizes the angle between magnetic field and chromospheric fibrils
only in these nonzero points. This local correction does, however, influence
the magnetogram globally, because the $L_1$ and $L_2$ term contain global
measures and the $L_4$ terms couples neighbouring points. As a consequence
the preprocessing result is different from classical preprocessing, even if
the $L_5$ term is nonzero only for a limited number of pixel.

For observational data
the weighting $w_4$ (last row in the tables, areas with less than $10\%$
ignored; see also \ref{fig_halpha} lower right panel)  seems to be more
realistic. In this case the overall average angle is not better than
for classical preprocessing, but is different by only about $3^\circ$
when preferential weighting is given to the more important strong field
regions.

The ultimate test regarding the success of our extended preprocessing scheme
is to use the preprocessed field as boundary conditions for a nonlinear
force-free coronal magnetic field extrapolation. The results are presented in
Table \ref{tab:extraparms}, row 3 for classical preprocessing and rows 4-7 for
H$\alpha$ preprocessing. We find that all preprocessed fields provide much
better results than using the unprocessed data. For classical preprocessing we
get the magnetic energy $\epsilon_{\rm mag}$ correct with an error of $3 \%$ (for
unprocessed data we got an error of $35 \%$).  Taking the H$\alpha$
information into account improves the result and the magnetic energy is
computed with an accuracy of $1\%$ or better, even for the cases where we used
chromospheric information only in parts of the entire regions.

\begin{table}
\begin{flushleft}
%\begin{tabular}{|c|c|c|c|c|c|c|c|c|r|r|}
\begin{tabular}{|ccc|cccc|cc|rr|}
\hline
Data & \multicolumn{2}{c|}{Weights} &
\multicolumn{4}{c|}{L$_{\rm prep} \times 10^{-6}$} & \multicolumn{2}{c|}{Aly criteria} & $\phi_{\rm ave}$ &  $\phi_{\rm ave,w}$ \\
~ & $\mu_{5}$ & $w$ & $L_{12}$ & $L_{3}$  &  $L_{4}$ &  $L_{5}$ &
$\varepsilon_{\rm force}$ & $\varepsilon_{\rm torque}$ & \small{[deg]} & \small{[deg]~~}\\
\hline
& & & & & & & & & &\\
$B_{\rm ch}$ & ~-~ & ~-~ & 452.137 & 3.57 & 0.18 & 0.00 & 0.0171 & 0.0203 & -- & --\\
$B_{\rm np}$ & ~-~ & ~-~ & 338287. & 0.00 & 4.45 & 0.49 & 0.4138 & 0.5797 & 19.2 & 18.9\\
$B_{\rm cp}$ & ~-~ & ~-~ & 0.06658 & 2.30 & 0.18 & 0.21 & 0.0003 & 0.0001 & 10.1 & 8.8 \\
& & & & & & & & & &\\
$B_{\rm H\alpha p}$ & 1.525 & w$_1$ & 33.37 & 2.45 & 0.17 & 0.0007 & 0.0062 & 0.0011 & 1.1& 0.4 \\
$B_{\rm H\alpha p}$ & 1.765 & w$_2$ & 31.70 & 2.47 & 0.15 & 0.0171 & 0.0060 & 0.0012 & 7.3& 2.0 \\
$B_{\rm H\alpha p}$ & 1.880 & w$_3$ & 29.10 & 2.41 & 0.15 & 0.1355 & 0.0058 & 0.0012 & 10.8& 6.8 \\
$B_{\rm H\alpha p}$ & 2.115 & w$_4$ & 32.16 & 2.41 & 0.17 & 0.0531 & 0.0060 & 0.0012 & 10.4& 3.1 \\
%& & & & & & & & & & \\
\hline
\end{tabular}

\end{flushleft}
\caption{Results of the various preprocessing schemes: the model
chromosphere and photosphere (first two rows), classical preprocessing (third
row), and the H$\alpha$ preprocessing cases (last four rows).  Column 1
identifies the data set, columns 2 and 3 the value of
$\mu_5$ and the weighting scheme used for the H$\alpha$ preprocessing cases.
Columns 4-7 provide the value of the functionals $L_{12}=L_1+L_2$, $L_3, L_4,
L_5$ as defined in Equations (\ref{defL_1})-(\ref{defL_4}) and (\ref{defL5}),
respectively.  In columns 8 and 9 we show how well the force-free and
torque-free consistency criteria ($\varepsilon_{\rm force}, \varepsilon_{\rm torque}$)
as defined in Equations (\ref{eps_force}) and (\ref{eps_torque}) are fulfilled. The
last two columns contain the averaged angle ($\phi_{\rm ave}=\langle \phi(x,y)
\rangle $) of the field with the model chromospheric data and a magnetic field
weighted average angle ($\phi_{\rm ave, w}=\frac{\langle B^2 \, \phi(x,y)
\rangle }{\langle B^2 \rangle }$) with $\phi(x,y)$ as defined in
Equation (\ref{angle_phi}).}
\label{tab:testparms2}

\end{table}

\begin{table}
\begin{flushleft}
%\begin{tabular}{|l|l|l|r|c|c|l|l|c|}
\begin{tabular}{|lll|rccllc|}
\hline
%Data&% \multicolumn{2}{|c|}{Weights} & ~ & \multicolumn{5}{c|}{Metrics}
 %\\ ~
 Data & $\mu_{5}$ & $w$ & $L$ & $C_{\rm vec}$ & $C_{\rm cs}$ & $E'_{\rm n}$& $E'_{\rm m}$ &
 $\epsilon_{\rm mag}$ \\
 \hline ~& ~ & ~ & ~ & ~ & ~ & ~ & ~ & ~ \\
 $B_{\rm ch}$ & ~-~
 & ~-~ & 0.53 & 1.00 & 1.00 & 1.00 & 1.00 & 1.00\\ $B_{\rm np}$ & ~-~ & ~-~ &46.03
 & 0.91 & 0.99 & 0.69 & 0.85 & 0.65\\ $B_{\rm cp}$ & ~-~ & ~-~ & 5.99 & 0.97 &
 0.99 & 0.80 & 0.85 & 0.97\\ ~& ~ & ~ & ~ & ~ & ~ & ~ & ~ & ~\\ $B_{\rm H\alpha
 p}$ & 1.525 & w$_1$ & 3.45 & 0.97 & 1.00 & 0.81 & 0.85 & 1.01\\ $B_{\rm H\alpha
 p}$ & 1.765 & w$_2$ & 2.37 & 0.97 & 1.00 & 0.81 & 0.86 & 1.00\\ $B_{\rm H\alpha
 p}$ & 1.880 & w$_3$ & 2.31 & 0.97 & 1.00 & 0.81 & 0.85 & 0.99\\ $B_{\rm H\alpha
 p}$ & 2.115 & w$_4$ & 3.20 & 0.97 & 1.00 & 0.81 & 0.85 & 1.00\\ \hline
\end{tabular}
\end{flushleft}
\caption{Results of the nonlinear force-free field extrapolations in a 3D box
$(320 \times 320 \times 256)$.  The rows are the same as in Table
\ref{tab:testparms2}.  The first three columns identify the preprocessing
scheme, the value of $\mu_5$, and the weighting scheme as in Table
\ref{tab:testparms2}.  The fourth column contains the functional $L$, as
defined in Equation (\ref{defL1}), which tells us how well the force-free and
solenoidal conditions are fulfilled in the computational box.  In columns 5-9
we compare the extrapolated 3D magnetic field with the reference solution and
use different quantitative comparison metrics: the vector correlation
($C_{\rm vec}$), the Cauchy-Schwarz metric ($C_{\rm cs}$), the complement of the
normalized vector error (E$'_{\rm n}$), the complement of the mean vector error
($E'_{\rm m}$), and the total magnetic energy normalized to the reference field
($\epsilon_{\rm mag}$) as defined in Equations (\ref{equ:cvec})-(\ref{equ:ep}),
respectively. Perfect agreement for any of these comparison metrics
  is unity. }
\label{tab:extraparms}
\end{table}

\section{Conclusions and Outlook}
\label{conlusions}
Within this work we developed an improved algorithm for the preprocessing of
photospheric vector magnetograms for the purpose of getting suitable boundary
conditions for nonlinear force-free extrapolations. We extended the
preprocessing routine developed by \inlinecite{wiegelmann:etal06}, which is referred
to here as ``classical preprocessing''. The main motivation for this work is
related to the fact that active-region coronal magnetic fields are
force-free due to the low $\beta$ coronal plasma, but the magnetic
field vector can be measured with high accuracy only on the photosphere, where
the plasma $\beta$ is about unity and non-magnetic forces cannot be
ignored. Our original (``classical'') preprocessing removes these
non-magnetic forces and makes the field compatible with the force-free
assumption leading to more chromospheric-like configurations.  In
this study, we have found that by taking direct chromospheric observations
into account (such as by using fibrils seen in H$\alpha$ images), the
preprocessing is improved beyond the classical scheme.  This improved scheme
includes a term which minimizes the angle between the preprocessed
magnetic field and the fibrils. We tested our method with the help of a model
active region developed by
%\inlinecite{vanballegooijen:etal07}
van Ballegooijen {\it et al.} (2007), which includes the
forced photospheric and force-free chromospheric and coronal layers.
This model has been used by
%\inlinecite{metcalf:etal07}
Metcalf {\it et al.} (2007)
for an inter-comparison of
nonlinear force-free extrapolation codes. The comparison revealed that the
model coronal magnetic field was reconstructed very well if chromospheric
magnetic fields have been used as input, but in contrast the
reconstructed fields compared poorly when unprocessed model-photospheric data
were used. Classical preprocessing significantly improves the result, but the
H$\alpha$ preprocessing developed in this paper is even better as the main
features of the model corona are reconstructed with high accuracy.
 Our extended preprocessing tool provides a fair estimate of the chromospheric
magnetic field, which is used as boundary condition for computing the
nonlinear force-free coronal magnetic field.
In
particular, the magnetic energy in the force-free domain above
the chromosphere agrees with the model corona within $1\%$, even if only
strong-field regions of the model chromosphere, where the fibrils can be
identified with highest accuracy, influence the final solution.  From these
tests we conclude that our improved preprocessing routine is a useful tool for
providing suitable boundary conditions for the computation of coronal magnetic
fields from measured photospheric vector magnetograms as provided for example
from {\it Hinode}. The combination of preprocessing and nonlinear force-free field
extrapolations seem likely to provide accurate computation of the
magnetic field in the corona.

We will still not get the magnetic field structure in the relative thin layer
between/in the photosphere and the chromosphere correct, because here
non-magnetic forces cannot be neglected due to the finite $\beta$
plasma. Although this layer is vertically thin ({\it e.g.}, 2 vertical grid points
in the
%\inlinecite{vanballegooijen:etal07}
van Ballegooijen {\it et al.} (2007)
model compared to 256 vertical grid
points in the corona) it contains a significant part of the total magnetic
energy of the entire domain, see
%\inlinecite{metcalf:etal07}
Metcalf {\it et al.} (2007).  Unfortunately, this
part of the energy cannot be recovered by force-free extrapolations, because
the region is non-force-free.
Our improved preprocessing routine includes chromospheric information
and therefore provides us with a closer approximation of the chromospheric
magnetic field.  This leads to more accurate estimates of the total magnetic
energy in the corona.

A further improvement of the
preprocessing routine could be done with the help of additional observations,
{\it e.g.} the line-of-sight chromospheric field, as planned for SOLIS. One could
include these measurement directly in the $L_3$-term (\ref{defL_3}) either as
the only information or in some weighted combination with the photospheric
field measurement.  An investigation of the true 3D-structure of the thin
non-force-free layer between photosphere and chromosphere requires further
research. First steps towards non-force-free magnetohydrostatic extrapolation
codes \cite{wiegelmann:etal06b} might help to reveal the secrets of this
layer. Non force-free magnetic field extrapolations will require additional
observational constraints, because the magnetic field, the plasma
density and pressure must be computed self-consistently in one model.

%%%%%%%%%%%%%%%%%%%%%%%%%%%%%%%%%%%%%%%%%%%%%%%%%%%%%%%%%%%%%%%%%%%%%%%%%%%

\begin{acks}
    The work of T. Wiegelmann was
  supported by DLR-grant 50 OC 0501 and J.K. Thalmann got financial support by
  DFG-grant WI 3211/1-1.  M. DeRosa, T. Metcalf, and C.  Schrijver were
  supported by Lockheed Martin Independent Research funds.  We acknowledge
  stimulating discussions during the fourth NLFFF-consortium meeting in June,
  2007 in Paris.
\end{acks}

\appendix
\section{Extrapolation of Nonlinear Force-free Coronal Magnetic Fields}
\label{appendixA}
We briefly summarize our nonlinear force-free extrapolation code here, which
has been used to compute the 3D magnetic fields.  We solve the force-free
Equations (\ref{divB}) and (\ref{jxb}) by optimizing (minimizing) the
following functional:

\begin{equation}
L=\int_{V} \; \left[w_a \; B^{-2} \, |(\nabla \times {\bf B}) \times {\bf
    B}|^2 +w_b \; |\nabla \cdot {\bf B}|^2\right] \; d^3x \label{defL1},
\end{equation}
where $w_a(x,y,z)$ and $w_b(x,y,z)$ are weighting functions. It is obvious
that (for $w_a, w_b >0$) the force-free equations (\ref{divB}) and (\ref{jxb}) are
fulfilled when $L$ is zero. The optimization method was proposed by
%\cite{Wheatland:etal00}
Wheatland, Sturrock, and Roumeliotis (2000)
and further developed in
%\cite{wiegelmann:etal03}
Wiegelmann and Neukirch (2003).
Here we use the implementation of
%\cite{wiegelmann04}
Wiegelmann (2004)
which has been applied to data in
%\cite{wiegelmann:etal05b}
Wiegelmann et al. (2005).  In this article, we used a recent update
including of our code that included a multi-scale approach (see
%\cite{metcalf:etal07}
Metcalf et al. (2007)
for details). This version of the optimization code was
also used with the (same as in this paper) model-chromospheric, photospheric
and classical preprocessed photospheric magnetic field vector as part of an
inter-code-comparison in \cite{metcalf:etal07}.  For alternative methods to
solve the force-free Equations (\ref{divB}) and (\ref{jxb}) see the review
papers by \cite{sakurai89,aly89,amari:etal97,mcclymont:etal97,wiegelmann07a}
and references therein.

\section{Metrics to Compare a 3D Coronal Magnetic Field with a Reference Solution.}
\label{appendixB}
In order to quantify the degree of agreement between the extrapolated vector
fields of the input model field (\textbf{B}, {\it i.e.}, the extrapolated
chromospheric (reference) field) and the nonlinear force-free solutions
(\textbf{b}, {\it i.e.}, the extrapolated preprocessed photospheric field) that are
specified on identical sets of grid points, we use five metrics in table
\ref{tab:extraparms} that compare either local characteristics or the global
energy content in addition to the force and divergence integrals. These
measures have been developed in
%\cite{schrijver:etal06}
Schrijver et al. (2006)
and subsequently been
used to evaluate the quality of force-free and non-force-free extrapolation
codes
\cite{amari:etal06,wiegelmann:etal06a,wiegelmann:etal06b,song:etal06,wiegelmann07,metcalf:etal07}.

 The vector correlation metric has been defined as

      \begin{equation}
         C_{\rm vec} =   \frac{\sum_{i} \mathbf{B}_i \mathbf{b}_i}
                {\sqrt{ \sum_i |\mathbf{B}_i|^2 \sum_i |\mathbf{b}_i|^2}},
                \label{equ:cvec}
      \end{equation}
where $\textbf{B}_i$ and $\textbf{b}_i$ are the vectors at each point
      $i$. One finds that $C_{\rm vec} = 1$ if the vector fields are identical and
      $C_{\rm vec} = 0$ if $\textbf{B}_i \perp \textbf{b}_i$.

   The Cauchy-Schwarz metric is based on the homonymous inequality
      ($|\bf{a} \cdot \bf{c}| \leq |\bf{a}||\bf{c}|$ for any two
      vectors $\textbf{a}$ and $\textbf{c})$

      \begin{equation}
       C_{\rm cs}  =     \frac{1}{M}
            \sum_{i} \frac{\bf{B}_i \cdot \bf{b}_i}
                          {|\bf{B}_i||\bf{b}_i|} \equiv
            \frac{1}{M} \sum_{i} \cos \theta_i,
            \label{equ:ccs}
      \end{equation}
where $M$ is the total number of vectors in the volume, and $\theta_i$
      the angle between \textbf{B} and \textbf{b} at point $i$ . It is entirely
      a measure of the angular differences of the vector
      fields, i.~e.~$C_{\rm cs} = 1$ if \textbf{B} $\parallel$ \textbf{b},
      $C_{\rm vec} = -1$ if they are anti-parallel, and
      $C_{\rm vec} = 0$ if $\textbf{B}_i \perp \textbf{b}_i$ at each point.

       The normalized vector error is defined as

      \begin{equation}
       E_{\rm n} =      \frac{\sum_{i} |\bf{b}_i - \bf{B}_i|}
                 {\sum_{i} |\bf{B}_i|}.
                 \label{equ:en}
      \end{equation}

   The mean vector error averages over relative differences  and is given by

      \begin{equation}\label{equ:em}
          E_{\rm m} =   \frac{1}{M}
            \sum_{i} \frac{|\bf{b}_i - \bf{B}_i|}
                         {|\bf{B}_i|}.
      \end{equation}

      Unlike the first two metrics, perfect agreement of the two vector fields
      results in $E_{\rm m} = E_{\rm n} = 0$. For an easier comparison with the others,
      we list $E'_{\rm m,n} \equiv 1 - E_{\rm m,n}$, so that all
      measures reach unity for a perfect match.

  To estimate how well the models rates the energy content of the field,
      we use the total magnetic energy of \textbf{b}, normalized to the total
      magnetic energy of \textbf{B}, namely

      \begin{equation}
         \epsilon_{\rm mag} =     \frac {\sum_{i} |\bf{b}_i|^2}
                  {\sum_{i} |\bf{B}_i|^2}.
                  \label{equ:ep}
      \end{equation}

%%% BIBLIOGRAPHY %%%%%%%%%%%%%%%%%%%%%%%%%%%%%%%%%%%%%%%%%%%%%%%%%%%%%%%%%%%

     % format of references provided by the journal (.bst)
%\bibliographystyle{spr-mp-sola}

     % name your Bibtex file containing your references (.bib)
%\bibliography{SOLA_bibliography_example}
%\bibliography{tw}
     % Checking: look if the file containing the ``\bibitem'' exits
     %           so check if the .bbl file exist (bibTeX compilation)

\end{article}
\end{document}